\documentclass[sigconf]{acmart}
\pdfoutput = 1
\usepackage[utf8]{inputenc}

\usepackage{enumerate}
\usepackage{siunitx}
\usepackage{wrapfig}
\usepackage{xspace}

\newcommand{\figref}[1]{\figurename~\ref{#1}}

\renewcommand{\paragraph}[1]{\smallskip\par\emph{#1}.}

\newenvironment{authorguide}{\begin{quotation}\footnotesize\itshape}{\end{quotation}}

\newcounter{ExperimentStep}
\newcommand{\expstep}{\stepcounter{ExperimentStep}\paragraph{\alph{ExperimentStep}}\xspace }

\newcounter{AnalysisStep}
\newcommand{\analysisstep}{\stepcounter{AnalysisStep}\paragraph{\alph{AnalysisStep}}\xspace }

\newif\ifwrapfigs \wrapfigsfalse

\copyrightyear{2020}
\acmYear{2020}
\setcopyright{licensedothergov}
\acmConference[MSR '20]{17th International Conference on Mining Software Repositories}{October 5--6, 2020}{Seoul, Republic of Korea}
\acmBooktitle{17th International Conference on Mining Software Repositories (MSR '20), October 5--6, 2020, Seoul, Republic of Korea}
\acmPrice{15.00}
\acmDOI{10.1145/3379597.3387506}
\acmISBN{978-1-4503-7517-7/20/05}

\title[The Intrinsic Structure of Public Software Development History]{Determining the Intrinsic Structure of\\ Public Software Development History}

\author{Antoine Pietri}
\email{antoine.pietri@inria.fr}
\orcid{0000-0003-4052-4469}
\affiliation{\institution{Inria}
\city{Paris}
  \state{France}
}

\author{Guillaume Rousseau}
\email{guillaume.rousseau@univ-paris-diderot.fr}
\orcid{0000-0003-3583-995X}
\affiliation{\institution{Université de Paris and Inria}
\city{Paris}
  \state{France}
}

\author{Stefano Zacchiroli}
\email{zack@irif.fr}
\orcid{0000-0002-4576-136X}
\affiliation{\institution{Université de Paris and Inria}
\city{Paris}
  \state{France}
}

\begin{document}

\begin{abstract}

  \paragraph{Background}
  Collaborative software development has produced a wealth of version control
  system (VCS) data that can now be analyzed in full. Little is known about the
  intrinsic structure of the entire corpus of publicly available VCS as an
  interconnected graph. Understanding its structure is needed to determine the
  best approach to analyze it in full and to avoid methodological pitfalls when
  doing so.

  \paragraph{Objective}
  We intend to determine the most salient network topology properties of public
  software development history as captured by VCS. We will explore: degree
  distributions, determining whether they are scale-free or not; distribution
  of connect component sizes; distribution of shortest path lengths.

  \paragraph{Method}
  We will use Software Heritage---which is the largest corpus of public VCS
  data---compress it using webgraph compression techniques, and analyze it
  in-memory using classic graph algorithms. Analyses will be performed both on
  the full graph and on relevant subgraphs.

  \paragraph{Limitations}
  The study is exploratory in nature; as such no hypotheses on the findings is
  stated at this time. Chosen graph algorithms are expected to scale to the
  corpus size, but it will need to be confirmed experimentally. External
  validity will depend on how representative Software Heritage is of the
  software commons.

\end{abstract}


\maketitle

\section{Introduction}
\label{sec:intro}

The rise in popularity of Free/Open Source Software (FOSS) and collaborative
development platforms~\cite{kalliamvakou2014promises} over the past decades has
made publicly available a wealth of software source code artifacts (source code
files, commits with all associated metadata, tagged released, etc.), which have
in turn benefited empirical software engineering (ESE) and mining software
repository (MSR) research. Version control systems (VCS) in particular have
been frequently analyzed~\cite{kagdi2007msrsurvey} due to the rich view they
provide on software evolution and their ease of exploitation since the advent
of distributed version control systems.

Only very recently systematic initiatives~\cite{swhipres2017, swhcacm2018,
  mockus2019woc} have been established to gather as much public VCS data as
possible in a single logical place, enabling ESE research at the scale of,
ideally, \emph{all} publicly available source code artifacts---i.e., our entire
\emph{software commons}~\cite{kranich2008information}.
In this paper we describe how we will conduct the \emph{first systematic
  exploratory study on the intrinsic structure of source code artifacts stored
  in publicly available version control systems}.

\ifwrapfigs
\begin{wrapfigure}{r}{0.55\linewidth}
  \vspace{-5mm}
\else
\begin{figure}
\fi
  \centering
  \includegraphics[width=\ifwrapfigs\linewidth\else 0.6\linewidth\fi,trim=1cm 1cm 1cm 1cm]{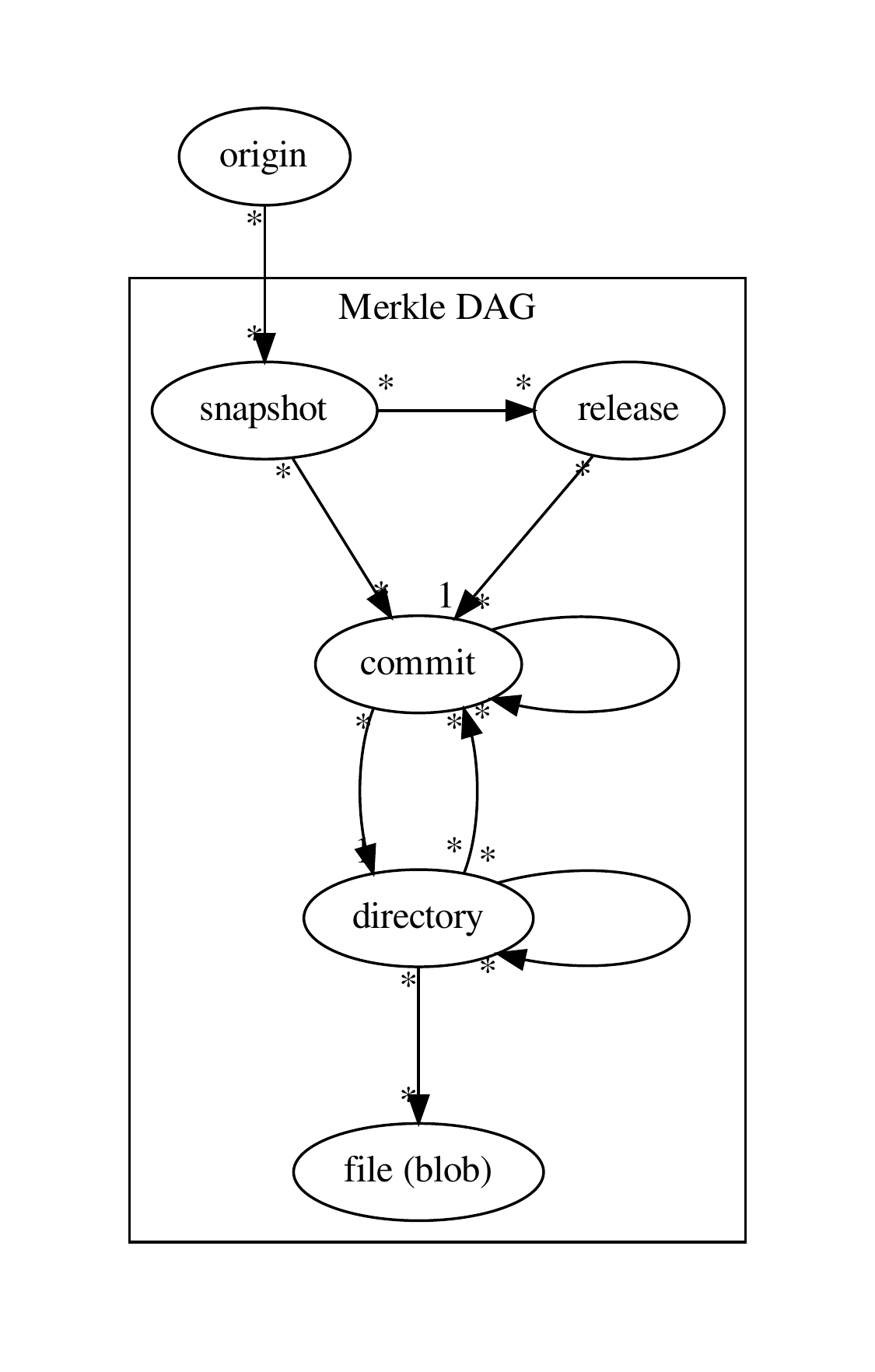}
  \caption{Data model: a Merkle DAG linking together source code artifacts
    commonly found in version control systems.}\label{fig:data-model}
\ifwrapfigs
\end{wrapfigure}
\else
\end{figure}
\fi

As corpus we will use Software Heritage and its dataset~\cite{swhcacm2018,
  swh-msr2019-dataset}, which is the largest and most diverse collection of
source code artifacts, spanning more than 5 billion unique source code files
and 1 billion unique commits, collected from more than 80 million software
projects.

The data model of our corpus is shown in \figref{fig:data-model}. It is a
Merkle DAG~\cite{Merkle}, fully deduplicated, linking together in a single
giant graph the development histories of all publicly available VCSs. Actual
source code files are represented as blob nodes, their grouping in source code
trees as directory nodes; commit nodes are linked together by ancestry,
supporting forks and N-ary merges; releases (or ``tags'') are also supported;
the full states of VCSs is recorded periodically (similarly to what the
Internet Archive wayback machine does for web pages) and associated to where it
has been observed (a software origin, identified by an URL). We refer the
reader to~\cite{swh-msr2019-dataset} for full details on the data model.

The study we will conduct will assess the most salient network topology
properties~\cite{barabasi2002networkstats} of the Software Heritage corpus as a
graph. Such a systematic analysis is still missing in the literature and is
needed for further empirical software engineering research for several reasons:

\paragraph{a) Determine the most appropriate large-scale analysis approach}
Most ``large-scale'' studies of VCSs fall short of the full body of publicly
available source code artifacts and either resort to random sampling or focus
on popular repositories. This is a potential source of bias, but is
understandable for practical reasons. To enable studies on the entire software
commons we need, in addition to platforms~\cite{swhcacm2018, mockus2019woc}, an
understanding of its intrinsic structure, to choose the most appropriate
large-scale analysis approach ~\cite{swh-provenance-tr, dyer2013boa} depending
on the study needs. For instance, if the graph is easy to partition into
loosely connected components, then a scale-out approach
with several compute nodes holding in memory graph quasi-partitions would be
best; if the graph is highly connected a scale-up approach relying on graph
compression~\cite{saner-2020-swh-graph} would be preferable. Similarly, knowing
that most nodes are part of a single gigantic connect component (CC) would help
in avoiding algorithmic approaches with high complexity on the size of the
largest CC.

\paragraph{b) Avoid methodological pitfalls}
The same understanding is needed to avoid making strong assumptions on what
constitutes ``typical'' VCS data. These pitfalls have been warned against since
the early days of GitHub~\cite{kalliamvakou2014promises}, but that have not
been quantified yet.  The extent to which repositories on popular forges
correspond to ``well behaved'' development repositories, as opposed to be
outliers that are not used for software development or are built just to test
the limits of hosting platforms or VCS technology is unknown. In our experience
GitHub alone contains repositories with very weird artifacts (commits with one
million parents or mimicking bitcoin mining in their IDs, the longest possible
paths, bogus timestamps~\cite{swh-provenance-tr}, etc.). How many statistically
relevant outliers of this kind exist is unknown and needs to be documented as
reference knowledge to help researchers in the interpretation of their
empirical findings.

\paragraph{c) Improve our understanding of our daily objects of study}
More generally, in empirical software engineering we are collectively studying
artifacts that naturally emerge from the human activity of software
development. As it is commonplace in other sciences (and most notably physics),
we want to study the intrinsic network properties of the development history of
our software commons just because the corpus exists, it is available, and it is
challenging to do so. The resulting findings might be \emph{also} practically
useful, but in spite of that we will obtain a more deep understanding of the
nature of objects we study daily than what is known today.

\section{Research Questions}
\label{sec:rq}

Specifically we will perform an exploratory study, with no predetermined
hypotheses, and answer the following research questions:
\newcommand{\RQdegrees}{RQ\ref{rq:degrees}\xspace}
\newcommand{\RQpaths}{RQ\ref{rq:paths}\xspace}
\newcommand{\RQccs}{RQ\ref{rq:ccs}\xspace}
\begin{enumerate}[\bfseries RQ1]

\item \label{rq:degrees}
What is the distribution of indegrees, outdegrees  and local clustering of the public VCS
  history graph? Which laws do they fit? \\
  How do such distributions vary across the different graph layers---file
  system layer (files + directories) v.~history layer (commits + releases)
  v.~origin layer.

\item \label{rq:ccs} What is the distribution of connected component sizes for
  the public VCS history graph?  How does it vary across graph layers?

\item \label{rq:paths} What is the distribution of shortest path lengths from
  roots to leaves in the recursive layers (commits and directories) of public
  VCS history graph.

\end{enumerate}

\section{Variables}
\label{sec:variables}

We define the following subsets of the starting corpus:
\begin{itemize}

\item \emph{full corpus}: the entire Software Heritage graph dataset

\item \emph{filesystem layer}: full corpus subset consisting of file and
  directory nodes only, and edges between them

\item \emph{history layer}: full corpus subset consisting of commit and
  releases only, and edges between them

\item \emph{commit layer}: subset of the history layer consisting of commit
  nodes only, and edges between them

\item \emph{hosting layer}: full corpus subset consisting of origins and
  snapshost nodes only, and edges between them

\end{itemize}

For each corpus we will measure the following variables:
\begin{itemize}

\item \emph{indegree distribution}: for each node the number of edges pointing
  to it

\item \emph{outdegree distribution}: for each node the number of edges starting
  from it

\item \emph{local undirected clustering distribution}: for each node the number 
    of edges between nodes pointing to or from it (= local clustering coefficient without dividing it by the number of possible egdes between its neighbors)

\item \emph{CC size distribution}: the size, in number of nodes, of each
  connected component in the underlying undirected graph of the input corpus

\item (for the filesystem and commit layers only) \emph{path length
    distribution}: distribution of the length of shortest paths between all
  root (indegree=0) nodes and all leave (outdegree=0) nodes
  
\end{itemize}

\section{Material}
\label{sec:material}
\label{sec:dataset}

\ifwrapfigs
\begin{wrapfigure}{r}{0.22\textwidth}
  \vspace{-5mm}
\else
\begin{table}
\fi
  \def\captiontext{Corpus size as a graph (release: 2018-09-25).}
  \ifwrapfigs\else\caption{\captiontext}\fi
  \centering \ifwrapfigs\else\small\fi
  \begin{tabular}{l|r}
    \multicolumn{2}{c}{\textbf{Nodes}} \\
    \hline\hline
    origins (ori)     & 85 M \\
    snapshots (snp)   & 57 M \\
    releases (rel)    & 9.9 M \\
    commits (cmt)     & 1.1 B \\
    directories (dir) & 4.4 B \\
    files             & 5.0 B \\
    \hline
                      & $\approx$\,11\,B
  \end{tabular}\ifwrapfigs\\\else\hspace{5mm}\fi
  \begin{tabular}{l|r}
    \multicolumn{2}{c}{\textbf{Edges}} \\
    \hline\hline
    ori$\to$snp   & 74 M \\
    snp$\to$cmt   & 616 M \\
    cmt$\to$cmt   & 1.2 B \\
    cmt$\to$dir   & 1.2 B \\
    dir$\to$dir   & 49 B \\
    dir$\to$file  & 113 B \\
    \hline
                & $\approx$\,165\,B
  \end{tabular}
  \ifwrapfigs\caption{\captiontext}\fi
  \label{fig:corpus-stats}
\ifwrapfigs
\end{wrapfigure}
\else
\end{table}
\fi

\paragraph{Dataset}
We intend to use the Software Heritage graph dataset~\cite{swh-msr2019-dataset}
as corpus for the planned experiments. The main reasons for this choice are
that: (1) it is the largest dataset about publicly software development history
(see \ifwrapfigs\figref{fig:corpus-stats}\else Table~\ref{fig:corpus-stats}\fi~
for its size characteristics as a graph); (2) it is available as open data in
various
formats,\footnote{\url{https://annex.softwareheritage.org/public/dataset/graph/},
  retrieved 2020-01-09} including a simple nodes/edges graph representation;
(3) it has been chosen as topic for the MSR 2020 Mining
Challenge,\footnote{\url{https://2020.msrconf.org/track/msr-2020-mining-challenge},
  retrieved 2020-01-09} hence we expect that by the time experiments will be
run the body of related work around it will be substantial.

\paragraph{Software}
We will use \texttt{swh-graph}~\cite{saner-2020-swh-graph} and
WebGraph~\cite{boldi-vigna-webgraph-1} to produce and exploit compressed graph
representations of the input corpus (see Section~\ref{sec:execplan} for
details). Statistical analyses will be performed using popular
SciPy~\cite{mckinney2012scipy} components (NumPy, Pandas, \ldots). All
developed custom code and derived data will be released as a complete
replication package.

\paragraph{Hardware}
We expect to be able to run all experiments on a single server equipped with 24
Intel Xeon 2.20 GHz CPUs and 750\,GB of RAM, which is already available to us.

\section{Execution Plan}
\label{sec:execplan}

We will follow the experiment protocol described below.

\expstep Retrieve the most recent version of the Software Heritage graph
dataset~\cite{swh-msr2019-dataset} available, falling back to the one available
at the time of writing (2018-09-25) if no newer releases have been published.

\expstep Compress the textual node/graph representation of the full dataset to
a compact representation of its structure (which ignores all node metadata
except node types) using \texttt{swh-graph}. Repeat for each subgraph of
interest (see Section~\ref{sec:variables} for details).  This step is expected
to take about a week for the full graph and proportionally less time for the
various subgraphs.

\expstep (\RQdegrees) Compute indegrees, outdegrees and local clustering 
for all nodes in the graph, for all relevant (sub)graphs, exploiting the
WebGraph~\cite{boldi-vigna-webgraph-1} API with custom Java code.

\expstep (\RQccs) Compute all connected components on all relevant (sub)graphs
using well known algorithms~\cite{hopcroft1973graphalgos} that can be
implemented with custom Java code as visits (BFS or DFS) on the WebGraph graph
representations. A full graph visit is expected to take a few hours, scaling
down linearly for subgraphs.~\cite{saner-2020-swh-graph}

\expstep (\RQpaths) For each subgraph of interest (e.g., the file system layer
composed only of files and directory nodes), for all root nodes (e.g., source
code root directories), create a shortest path spanning tree to all leaves
(e.g. file or directory nodes with no children). Then export all path
lengths. This can be implemented with custom Java code realizing Dijkstra's
algorithm on top of the WebGraph API. We have no precise estimate of how long
this step will take; it will take more than a single full graph visit, due to
paths that will be re-explored, but it should remain manageable. Thread-base
parallelization is an option in case we need to speed up this step.

\section{Analysis Plan}
\label{sec:analysisplan}

\subsection{Descriptive statistics}

We will analyze the raw data obtained from the execution plan according to the
following analysis protocol.

\analysisstep (\RQdegrees) We will plot indegree/outdegree/local clustering distributions as
histograms, one histogram per distribution and per (sub)graph (see
Section~\ref{sec:variables} for details).
\begin{itemize}
\item

  One figure per distribution and graph will be produced, showing both the raw
  histogram (i.e., for each degree/clustering value observed, the number of nodes with
  that value) as well as the cumulative distribution function (CDF) given by
  the number of nodes whose degree/clustering is greater than or equal to a given
  value. One or more scales among lin-lin, lin-log, log-lin, log-log scales
  will be used, depending on presentation needs.

\item Each distribution will be qualitatively discussed by comparing it with a
  generic distribution functions (power law, exponential law, Poisson law).

\item In case they arise, the presence of outliers or other statistical
  anomalies with respect to more regular regions of the distribution will be
  highlighted (e.g., change of law, change of slope, \ldots), and may lead to
  specific investigations to determine their nature.

\end{itemize}

\analysisstep (\RQdegrees) The nature of the tail of each distribution will be
systematically analyzed and discussed.
\begin{itemize}

\item The first observed criterion will be the amplitude of the range of degree
  values, expressed in decades.

\item Then, we will use the discrete maximum likelihood estimator (MLE) to
  determine the scaling parameter~\cite[Eq.~3.7]{clauset2009powerlaw}.
  $$\alpha(d_{min})=1+n\left[\sum_{i=1}^{n}\ln{d_i\over d_{min}}\right]^{-1}$$
  where $d_i, i=1,\dots,n$ are the observed degree values such that
  $d_i\ge d_{min}$.

  Without prejudging, nor speculating whether the distributions will match
  power laws or not, we limit ourselves to the first step of the methodology
  proposed in~\cite[Box~1]{clauset2009powerlaw}, using the above estimator
  which depends on an arbitrary degree threshold $d_{min}$, beyond which the
  behaviour of the distribution is ignored, like a probe through the whole
  range of degrees.

The plot is displayed in log-lin scale over the degree range.

  Lacking precise information on the results at this stage, as well as
  knowledge about the impact of possible atypical events in the observed
  distributions, it is premature to plan to implement the subsequent steps of
  the aforementioned methodology.

\end{itemize}

\analysisstep (\RQccs) We will display connected component size, focusing on
VCS origins that contain at least one commit, as a set of histograms.
\begin{itemize}

\item As above, both raw distribution functions and cumulative distributions
  will be displayed.

\item To characterize the aggregation process into connected components as
  nodes of different types, or at different depths, are added to the corpus:
  \begin{itemize}

  \item We will produce histograms showing the raw distribution and CDF, for
    each layer and depth within each layer.

  \item For each of them, we will produce a table summarizing the number of
    isolated origins and size of the largest connected component.

  \item We will quantitatively compare the distribution functions by displaying
    the Kolmogorv-Smirnov distance between them, weighted by connected
    component sizes expressed as the number of origin nodes they contain.

  \end{itemize}

\end{itemize}

\analysisstep (\RQpaths) We will display the distribution of root-leaf shortest
path lengths for the recursive layers of the public VCS corpus (commit and
filesystem layers), as a pair of histograms.
\begin{itemize}

\item As before, both raw and cumulative distribution functions will be
  displayed and discussed.

\end{itemize}

\subsection{Practical significance of the findings}

\RQdegrees's answers will provide information on how the public VCS graph
compares with other naturally occurring graphs in collaboration, such as the
social network graphs or the graph of the Web. While seemingly only of
theoretical significance, degree distributions and derived properties such as
graph density directly impact on the practical exploitability of graphs of this
scale. For instance graphs with fat-tailed degree distributions compress better
than others. We know from previous work~\cite{swh-provenance-tr,saner-2020-swh-graph} how well
the full VCS graph compress, but a systematic study of related properties for
the subgraphs we intend to address here is still missing.

Findings related to \RQccs will directly tell how to best approach full-scale
analyses of the various corpuses. If, for instance, it will turn out that 80\%
of the nodes are part of a gigantic component, than distributed approaches will
be generally difficult to implement, due to entanglement and the pseudo-random
nature of identifiers in Merkle DAG. Opposite considerations will apply for a
more uniform distribution of CC sizes.

Finally, \RQpaths's answers will tell what's the average path depths in the
recursive layers of the public VCS corpus, which is practically useful for
several analysis needs. The average commit length is a limiting factor in
analyses based on \texttt{git blame}, as we need to travel back development
history to attribute files (if not SLOC) contributions. Similar considerations
apply to average path lengths (in the filesystem layer), when we need to
attribute files or directories to originating commits, for analysis or
provenance tracking needs.

\section{Limitations}
\label{sec:limitations}

\paragraph{Exploratory nature of the study}
The study we propose is exploratory in nature, hence we state no hypotheses on
the findings at this stage. This is intended, as the intrinsic structure of the
public VCS corpus has never been characterized before at the extent we propose.

\paragraph{Algorithmic feasibility}
With its 10+\,B nodes and 160+\,B edges the corpus we plan to study is a
substantial graph for graph practitioners standards; not as big as the graph of
the Web, but significantly larger than most benchmarks used in the field. At
this scale, algorithms with super-linear complexity are generally considered
non practically applicable.

Our algorithmic approaches for \RQdegrees and \RQccs have linear complexity;
the approach for \RQpaths is super linear, but we are restricting it to
selected subgraphs and we expect it to exhibit enough sharing/caching to be
practically treatable. We do not expect it to be the case but if, due to either
algorithmic of technological considerations, any specific sub-experiment will
turn out to be not practically feasible, we will resort to uniform random
sampling.

\paragraph{External validity}
Due to organic crawling lag, Software Heritage does not capture the full extent
of publicly available VCSs. Hence we do not claim being able to characterize
the intrinsic structure of the entire history of publicly available software
development. The chosen dataset is nonetheless the best publicly accessible
option available today to researchers. It is also well representative of the
most popular development forge(s) in use today. As such we expect that the
network topology findings of this study will provide useful insights to
researchers and practitioners in the field.


\end{document}